# Dependence of Phoretic Propulsion of Sub-Micron Urease Motors on the Fraction of the Particle Covered by Catalyst.

*Roshan S Velluvakandy[1], Atul C Chaskar[1,2]*

[1] *National Centre for Nanoscience and Nanotechnology, University of Mumbai, India.*

[2] *Institute of Chemical Technology, Mumbai, India.*

**Abstract**:

Micro/Nano motors are micro to nano-sized particles and systems that are capable of motion in fluidic systems. Their motion can be a result of many mechanisms, one of the most well-known of which is electrolyte diffusiophoresis, in which asymmetric production of charged ions across the surface of the particle leads to phoresis. The focus of this work is to construct Urease motors with careful control of their Janus balance, i.e., to control the percentage of surface covered with the enzyme and study the effect of this asymmetry on the propulsion velocity of the particles. In that regard, we were successful in constructing particles with varied coating percentages. The observation made was that as the percentage coating approaches 50%, the propulsion velocity of the particle increases steadily. The trend may reverse at lower catalytic cap heights after reaching a certain threshold of enzyme activity and per cent coverage. However, the method used had limitations in constructing motors with lower catalytic cap heights. We aim to shed light on the design principles of efficient urease-driven phoretic motors, which could have significant implications for the development of next-generation nanoscale devices.

**Graphical Abstract**:

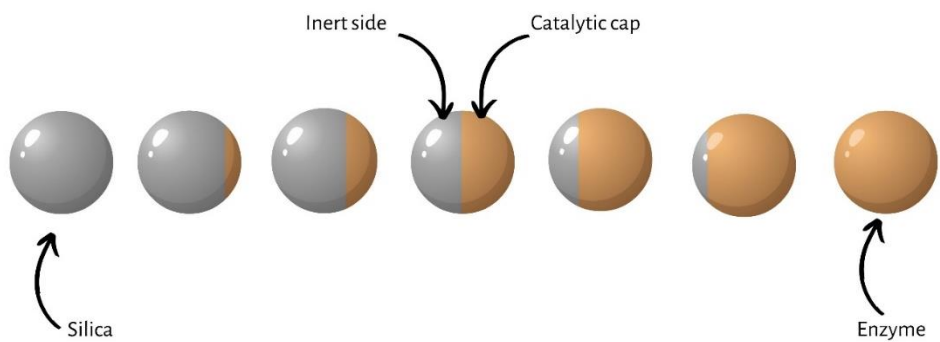

## Introduction

The field of nanotechnology has seen a surge in the development of synthetic micro- and nanomotors that can move autonomously through the use of surrounding chemical energy or external physical fields. Among the most promising of these micro-/nanomotors are Janus micro-/nanoparticles, which possess asymmetric surface chemistries, compositions, or compartments in a single structural unit. These particles can be propelled through self-generated chemical gradients resulting from asymmetric catalytic reactions. As such, Janus particles have shown promise in a wide range of applications, including drug delivery, molecular sensing, environmental remediation and in vivo imaging. However, the precise relationship between the fraction of the particle surface covered by the catalyst and its effect on the resulting phoretic propulsion remains unclear.

Many micro/nanomotors powered by inorganic catalysts have limitations of use in a biomedical scenario given the toxic or non-degradable nature of those catalysts. Furthermore, the fuel required to power enzyme-based motors is already present in the biological system. Given these advantages, there has been considerable development in the field of enzyme micro/nanomotors.

The most common enzymes used are urease[1–8], glucose oxidase[9–14] and catalase[15–22]. Other enzymes like ATPase[23], RNAse[24] and lipase[25] among others have also been reported.

The mechanism of propulsion of these different enzyme-based motors is however fairly varied. Motors based on urease for instance are based on ionic self-diffusiophoretic propulsion. Urease breaks down urea into ionic components, ammonium ions and carbonate ions, the diffusivities of which are different. $NH_4^+$= 1.957 x $10^{-5}$ $cm^2 s^{-1}$ and $CO_3^{2-}$= 0.923 x $10^{-5}$ $cm^2 s^{-1}$ [26] .This difference leads to the generation of ionic gradients which lead to the generation of local electric fields and fluid flow from regions of higher ionic concentrations to lower ionic concentrations and therefore leading to propulsion.

Some theoretical models have been described studying the influence of the active catalyst cover on the phoretic behaviour of the particle, particularly for neutral solute diffusiophoresis[27,28]. In terms of experimental demonstrations, there is only one report that studied the effect on the propulsion of varying coverage of platinum nanoparticles on dendritic silica particles. They observed that particles with 50 per cent coverage showed the highest propulsion velocities. Platinum and Peroxide based motors however are in principle different from Urease based motors[29]. Additionally, the exact mechanism of propulsion of this kind of platinum-peroxide motor remains controversial[30]. That being said, urease motors do not hold such uncertainty in their mechanistic underpinnings and would therefore be suitable to study the effect of variable catalyst coverage.

In this context, the present work investigates the dependence of phoretic propulsion of sub-micron urease motors on the fraction of the particle covered by the catalyst. By examining the relationship between the coverage of the catalyst and the resulting phoretic propulsion, we aim to shed light on the design principles of efficient urease-driven phoretic motors, which could have significant implications for the development of next-generation nanoscale devices.

## Results and Discussion

### Fabrication and Characterization of Urease motors

Submicron silica particles were synthesized using the Stöber method, also known as the sol-gel process. This method involves the hydrolysis and condensation of the silica precursor tetraethyl orthosilicate in an alcoholic solvent in the presence of water and a base catalyst. The resulting particles are chemically inert and have a narrow, monodisperse size distribution, making them ideal for use as the central core in enzyme motors.[31–33]

Scanning electron microscopy (SEM) images (Fig.1) showed that the synthesized particles were approximately 184±7.5 nm in size and therefore fairly monodisperse.

To modify the surface chemistry of the silica particles and enable the covalent attachment of enzymes, silane coupling agents were used. These agents, such as (3-aminopropyl) triethoxysilane (APTES) and (3-glycidyloxypropyl) triethoxysilane (GPTES), contain specific functional groups that can react with the surface of the silica particles. GPTES, in particular, contains an epoxide group that can react with -SH groups at neutral pH and with -$NH_2$ at slightly alkaline conditions.[34] The resulting covalent bond ensures a strong and stable attachment of the enzyme to the silica surface. Most urease motors described to date are synthesized using APTES modification followed by the use of a glutaraldehyde linker for enzyme attachment.[1,2,35,36] However, we chose to simplify the process by using GPTES for the surface modification of the silica particles.

DLS-based particle size measurement of GPTES treated particles shows a small increase in particle size as compared to Bare silica particles, the mean diameter increasing from 183 nm for Bare Silica particles to 190nm for GPTES grafted silica particles. We also monitor the zeta potential changes to confirm modification and subsequent conjugation. The Zeta Potential of Unmodified Bare Silica particles was recorded at -17 mV, GPTES grafting reduces the zeta potential to -24.9 mV and further Urease functionalization increases the Zeta potential to -19.5 mV. Urease functionalization also increases the mean particle size estimated by DLS to 203nm. The attachment of Urease to GPTES-modified silica particles is also evident in the FTIR spectrum of the enzyme-conjugated particles. FTIR spectrum of Unmodified Silica particles exhibits peaks around 800 $cm^{-1}$ and 1100 $cm^{-1}$ due to the symmetric and asymmetric vibrations of Si-O. The band in the 950 $cm^{-1}$ region is due to the asymmetric vibrations of Si-OH. The weak shoulder near 550$cm^{-1}$ is associated with the bending of the oxygen of the Si-O-Si bond. These peaks are characteristic of Silica. Characteristic peaks associated with Amides (~1650 $cm^{-1}$ and ~1540 $cm^{-1}$) can be observed in the Urease IR spectrum while that of Urease bound Silica particles exhibits clear characteristic peaks overlapping with those of Urease and Silica, thus confirming functionalization.

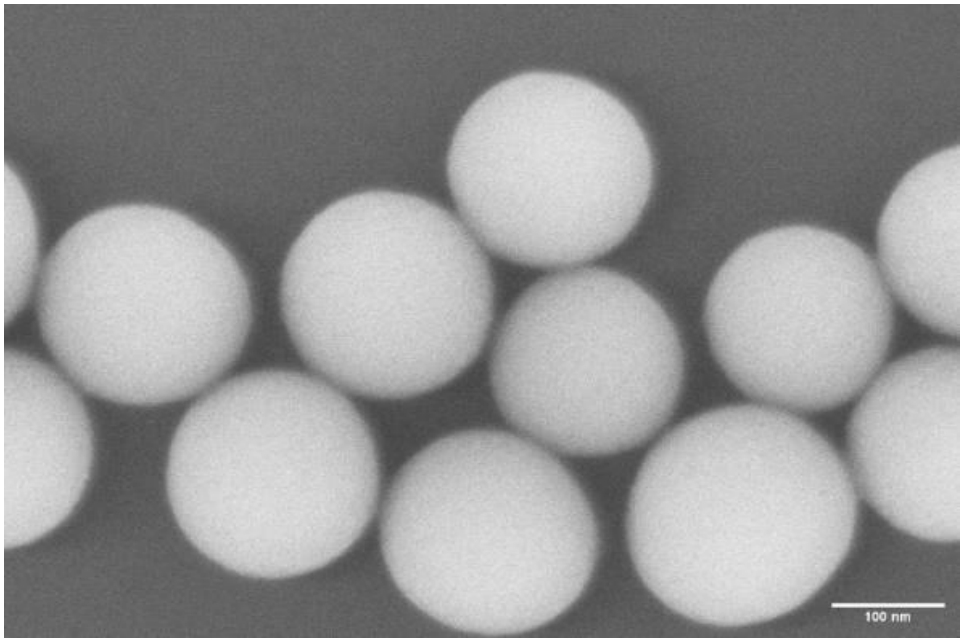

Fig.1: SEM micrograph of the synthesized Silica particles, Scale bar of 100nm

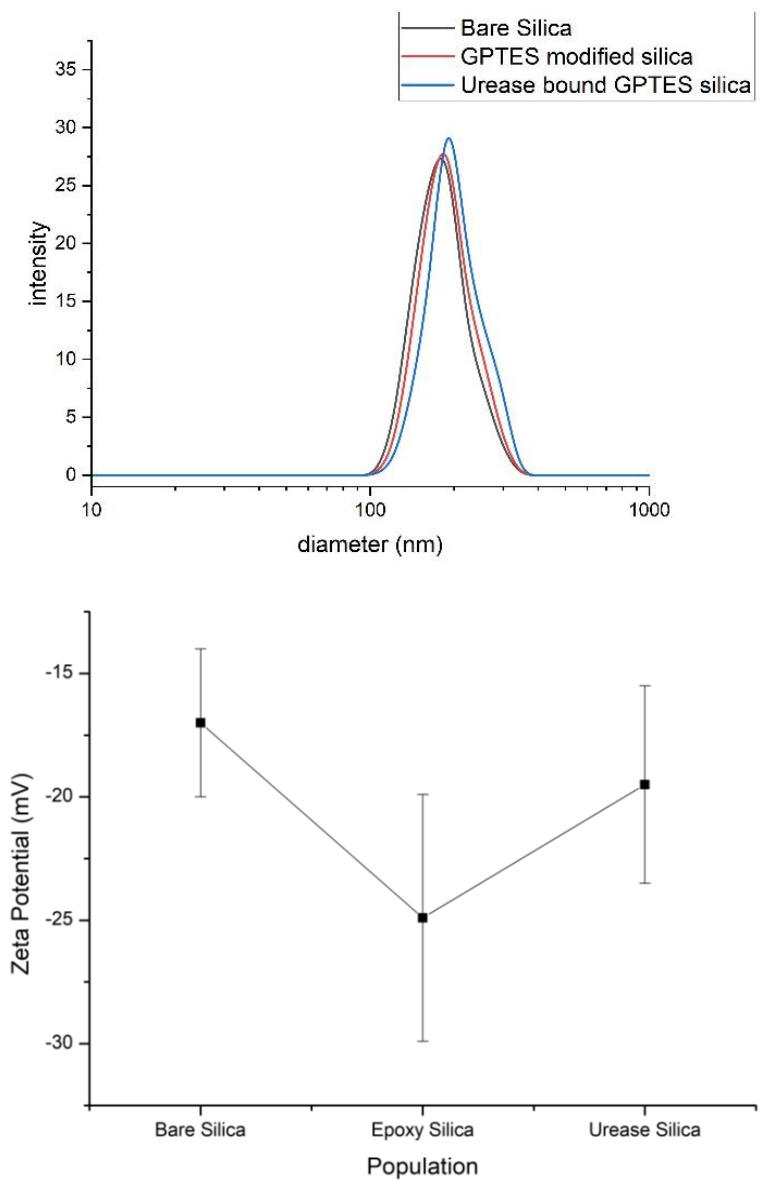

Fig.2: A) DLS particle size results show a marginal change in the size of the particles. B) Zeta potential measurements for unmodified silica, GPTES treated silica and Urease functionalized silica particles, error bar represents zeta deviation.

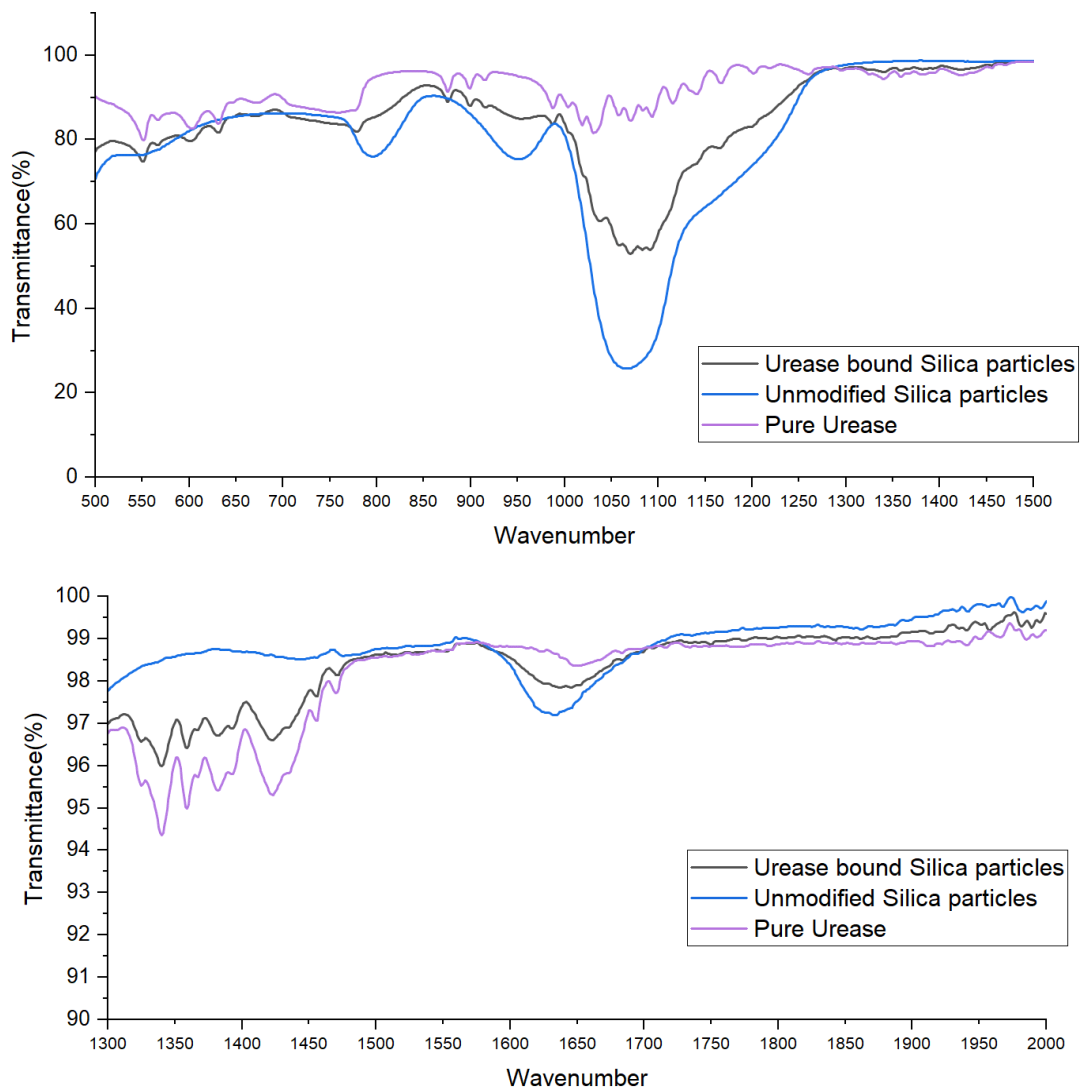

Figure 3: FTIR spectrum of Unmodified Silica particles, Pure enzyme Urease and Urease functionalized Silica Particles.

Janus particles, which have two distinct halves with different properties, were synthesized using a Pickering emulsion method described by Hong et.al in 2006[37]. This method involves vigorous mixing of molten paraffin wax and water, with silica particles acting as the stabilizing agent. As the emulsion cools, the wax solidifies, trapping the silica particles in place and masking one-half of their surface. When these Colloidosomes are exposed to enzymes, the unmasked half of the GPTES-modified silica is allowed to conjugate with enzymes. To control the balance of the Janus particles, the concentration of the surfactant Didodecyldimethylammonium bromide (DDAB) was adjusted. Increasing the concentration of DDAB led to an increase in the contact angle between the two phases, resulting in a shorter spherical cap exposed for enzyme conjugation.[38]

Colloidosomes synthesized with varying concentrations of DDAB from 15 mg/L to 60 mg/L were incubated with the enzyme urease to allow for conjugation of the enzyme to the particle surface. After incubation, the motors were separated from the wax by dissolving the colloidosomes in n-hexane. This resulted in the formation of urease-functionalized Janus particles. The different populations of synthesized particles were labelled as "A" for particles with 100 per cent of their surface functionalized with enzyme, "B" for particles synthesized with 15 mg/L DDAB, "C" for particles synthesized with 30 mg/L DDAB, "D" for particles synthesized with 15mg/L DDAB and "O" for non-functionalized control particles which were GPTES labelled silica particles.

To characterize how changing DDAD concentrations affect the Janus balance of the particles with Fluorescence associated cell cytometry (FACS), the colloidosomes were exposed to Fluorescein Cadaverine instead of Urease. These FITC-labelled particles were analysed with FACS to study how the fluorescence intensity varies with changing Janus balance.

As can be seen in the FITC intensity histogram, all 5 populations exhibit significantly different mean intensity peaks. However, there is also a not-so-insignificant overlap in the different populations. This leads to an observation that it may be challenging to exercise strict control over Janus balance following the method of wax colloidosomes. That being said, increasing surfactant concentration does have a clear influence on the Janus balance. As the concentration of DDAB increases the percentage of the sphere masked by wax also increases (as a result of increased depth of embedment) and this has an inverse relationship with the amount of FITC or Urease that can be functionalized on the surface. Another important factor to realise with this experiment is that FACS is generally unsuitable for the characterization of 200nm particles and therefore any attempt at a quantitative analysis would most likely be flawed. Therefore extrapolating the observations to exact percentage coverage would not be accurate.

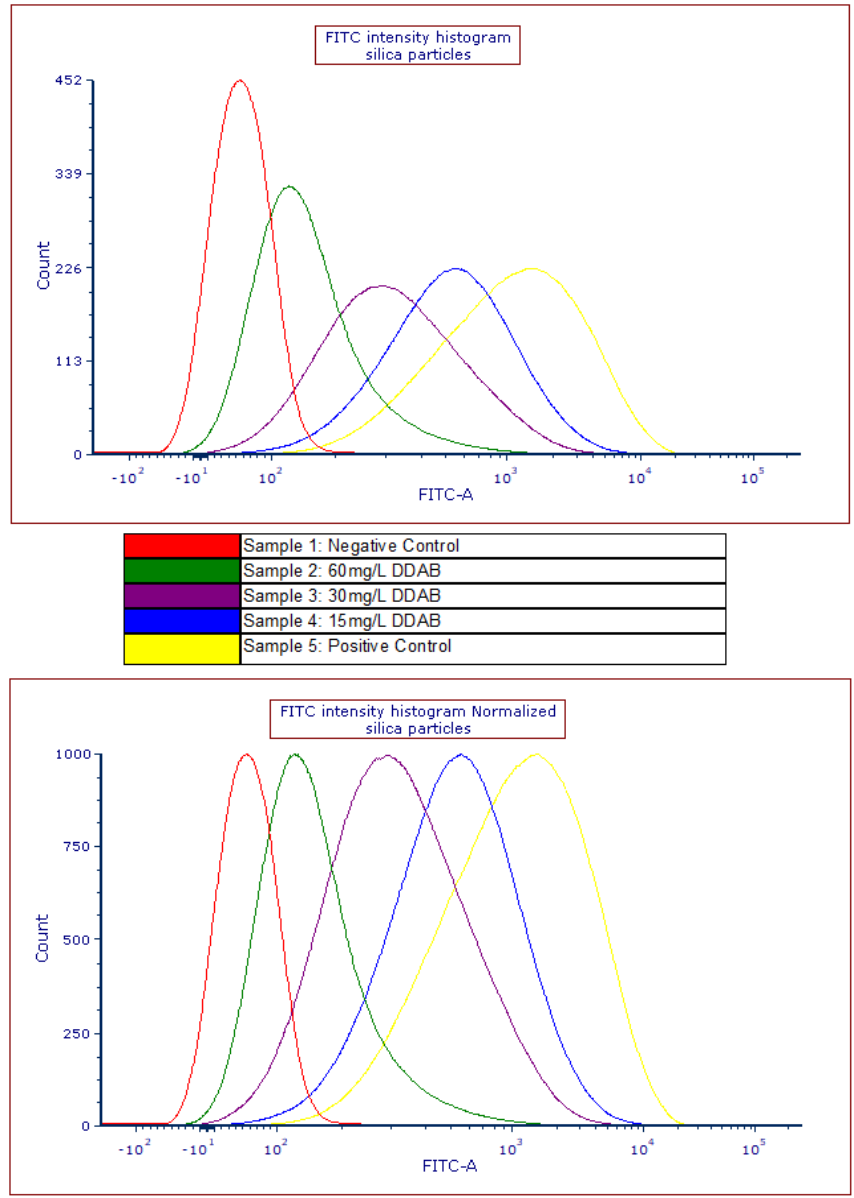

Fig.4: FACS histogram for FITC intensity, Gated for particles based on negative control. the x-axis in Hyper-Log.

**Enzymatic Assay of Urease labelled particles**

Enzymatic Activity of urease was carried out for silica particles of varying Janus balance labelled with urease and for particles coated throughout with urease. The titrimetric method was used for the determination of urease activity.

1ml of 110mM Urea prepared in Phosphate buffer pH 7.0 was added to blank and test tubes. To the test solution, 0.1ml of Urease labelled Silica suspension of an initial concentration of 11.0 mg/ml was added. The final concentration of silica, therefore, was 1mg/ml in a total volume of 1.10 ml. This mixture was set in an incubator at 25°C for 5 minutes after which 200 microliters of p-Nitrophenol indicator (40%w/v) was added to the solution. In the case of the blank, .1ml of Urease labelled Silica suspension of an initial concentration of 11.0

mg/ml was added after the incubation period and both tubes were quickly titrated against 1 micromolar HCL

(ΔV)HCl = (V)Test – (V)Blank

Units/ml of urease was defined as "a unit will liberate 1.0 micromole of NH3 from urea per minute at pH 7.0 at 25°C" and calculated as follows

((ΔV)HCl)/(5)(0.1)

The test was repeated 10 times and the average value of (ΔV)HCl was used for calculations

| Population | Units/ml enzyme |
|---|---|
| D | 1.3 |
| C | 1.4 |
| B | 1.8 |
| A | 2.2 |

**Diffusion Coefficient measurements**

The diffusion coefficient of the motors was measured with DLS in a Malvern Zetasizer Nano ZS. Urease-labelled Silica suspension was diluted to a final concentration of 0.1 mg/ml in specific concentrations of Urea. The reaction was let to reach 25°C in the Zetasizer for 120 seconds after which 3 readings of at least 13 runs each were measured. The average distribution was plotted.

To confirm motor synthesis, we first measure diffusion coefficients of the prepared population in increased concentration of the substrate urea. In all populations studied there was a clear increase in diffusion coefficients with an increase in substrate concentration.

As can be observed in Fig. 5, in the case of population "D", the Diffusion Coefficient of the motors increased from 2.4 µm$^2$/s in no Urea to 3.22 µm$^2$/s in 100mM Urea.

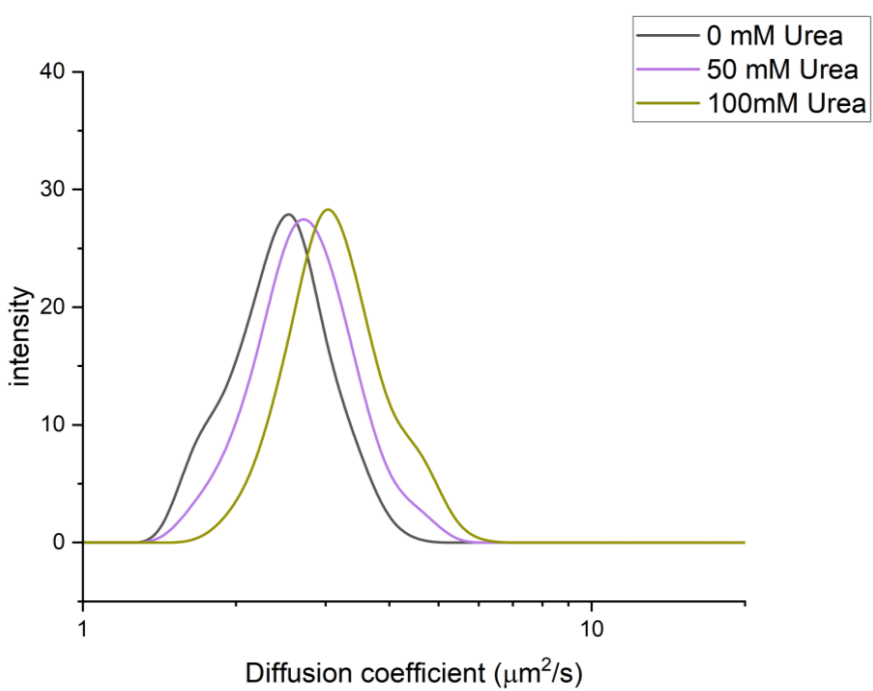

Fig. 5: Diffusion coefficient distribution of Sample "D" with varying concentrations of Urea.

Figure 6.A shows diffusion coefficients of all the different populations in 100mM Urea, as expected the lowest value of 2.48 µm$^2$/s was recorded for bare inactive particles. We observe a slightly higher diffusion coefficient of 2.59 µm$^2$/s for Population "A" which consists of particles with a complete surface coverage of Urease. Although in theory, symmetrically active particles do not lead to propulsion – in this case, we expect some asymmetry in the surface simply as a product of the functionalization method, 100% uniform coverage of Urease would be practically implausible and the asymmetry that results explains the observation. Also note this observation is in line with previous studies on the matter.[36] Next we observe a steady but substantial increase in the diffusion coefficients moving from populations "B" to "D" reaching a maximum of 3.20 µm$^2$/s. From the FACS data we expect that the height of the spherical active cap decreases as we go from "B" to "D" meaning as we approach lower per cent catalytic coverages the diffusion coefficients increase, we expect that after reaching a certain threshold in the relation of enzyme activity and per cent coverage this behaviour would reverse.

Figure 6. B shows the increase in diffusion coefficients after subtraction of control tests without Urea. We see the maximum increase at 0.75 µm$^2$/s for population "D" and the minimum at 0.11 µm$^2$/s for population "A".

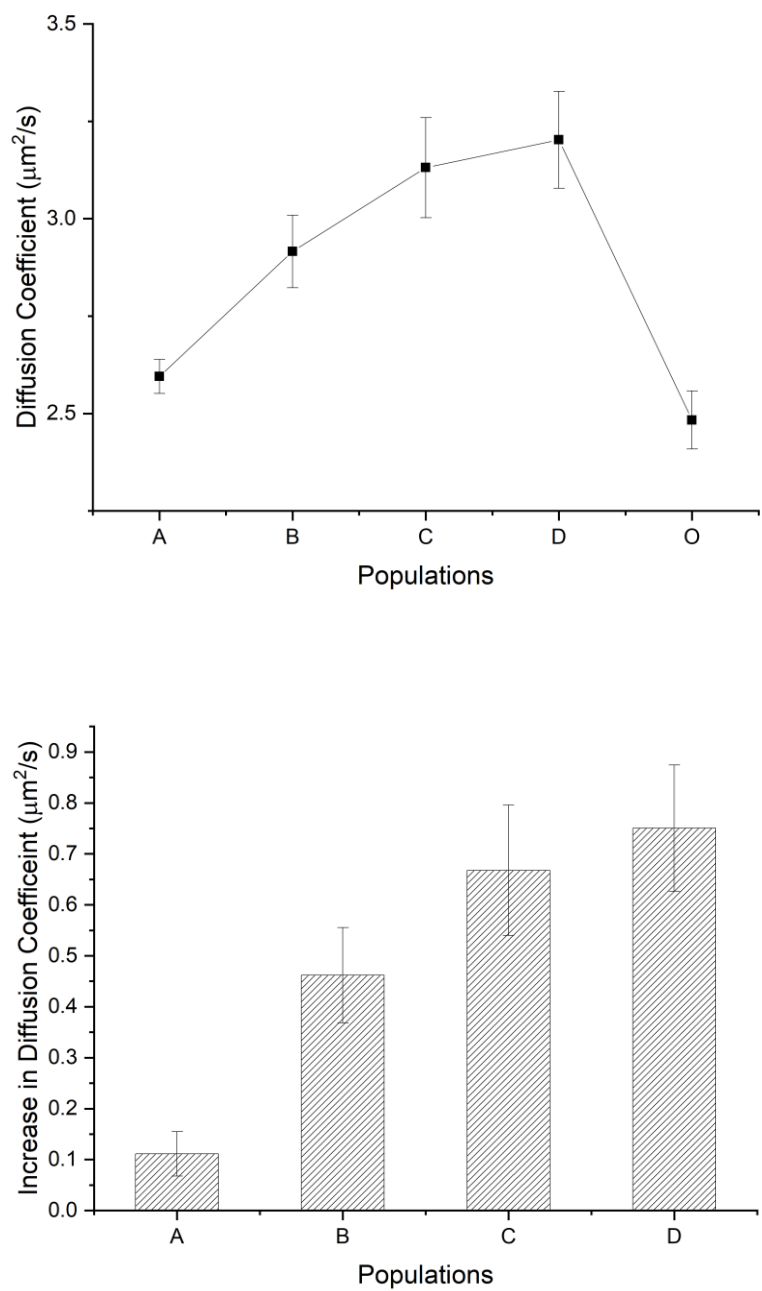

Fig. 6: A) Mean Diffusion coefficients of all the five populations in 100mM urea. The error bar represents the standard deviation. B) Mean increase in diffusion coefficients of the four populations are subtraction of diffusion coefficients in 0mM urea. The error bar represents the standard deviation.

Although we would like to construct motors with lower catalytic caps than the ones described in the work to study the threshold where the trend reverses, it is challenging given the limitations of the method used. The highest surfactant concentration where stable colloidosomes was found to be 60 mg/L. As we approach the Critical Micellar Concentration (CMC) or go past it - a phase inversion occurs where instead of forming an oil-in-water emulsion, a water-in-oil emulsion forms and the resultant particles are large and do not resemble colloidosomes. From previous work on controlling Janus Balance with surfactants, we expect that at 60mg/L we may either be close to 50 per cent coverage or approaching it.[38]

## Conclusion

In this study, submicron silica particles were synthesized using the Stöber method and surface-modified using silane coupling agents to enable covalent attachment of enzymes, such as urease. The use of (3-glycidyloxypropyl) triethoxysilane (GPTES) for surface modification was explored as a simpler alternative to the commonly used (3-aminopropyl) triethoxysilane (APTES) method. The particles were then used to create urease-functionalized Janus particles through a Pickering emulsion method. FACS analysis of fluorescently labelled particles found that in agreement with previous reports increasing the concentration of the surfactant didodecyldimethylammonium bromide (DDAB) led to a shorter spherical cap exposed for enzyme conjugation. However, there was a challenge in strictly controlling the Janus balance using this method. The study used Dynamic Light Scattering to measure the diffusion coefficient of motors labelled with urease and suspended in urea at varying concentrations. Results showed an increase in diffusion coefficients with increasing substrate concentration, and the trend was observed in all populations studied. The highest diffusion coefficient enhancement recorded was 0.75 µm$^2$/s in 100mM Urea. The study found that as the height of the catalytic cap decreases, diffusion coefficients increase, and the trend may reverse at lower catalytic cap heights after reaching a certain threshold of enzyme activity and per cent coverage. However, limitations in the method used, make it challenging to construct motors with lower catalytic caps heights. Synthesis of motors using alternate methods which transcend the limitations of lower catalytic cap heights may allow the study of the threshold of trend reversal and to examine the relationship between enzyme activity and catalytic coverage percentage in greater detail.

## Materials and Methods

### Materials

Tetraethyl orthosilane (TEOS, 98%), (3-Glycidyloxypropyl) triethoxysilane, (GPTES, 97.0%), Didodecyldimethylammonium bromide (DDAB, 98%) were purchased from Sigma Aldrich. Fluorescein Cadaverine (FITC, 98%) was purchased from Thermo Fischer. Paraffin Wax Pellets (Type 2 - 58-60°C), n-Hexane pure (99%), Urease ex. Jack Beans (200U/mg), Urea (99.5%), p-Nitrophenol (99.5%) 10X Phosphate Buffered Saline (PBS), Sodium Bicarbonate and Sodium Carbonate were purchased from SRLchem, India. Ammonia 25wt%, hydrogen peroxide 30%, Sulphuric acid and Hydrochloric acid were purchased from Molychem, India. Ethanol 99.9% was purchased from local distributors. Mili-Q Type 1 ultrapure water was used for all experiments.

### Methods

#### Synthesis of Monodisperse silica particles

Submicron Stober silica spheres were synthesized by methods described.[33]

A typical reaction consisted of 0.4M TEOS, 11M Water and 0.8M Ammonia in ethanol. In a 100ml reaction, first 35ml of Ethanol was mixed with 15.3 ml Water and 6ml Ammonia solution. The solution was mixed with the help of a magnetic stirrer. After 15 mins, 8.8ml of TEOS diluted with another 35ml of Ethanol was added with vigorous stirring. During the course of the reaction, the turbidity of the solution increases, going from a clear solution to a milky white suspension. The reaction is continued for 12 hours with moderate stirring at Room temperature. After 12 hours, the silica suspension Is subjected to centrifugation at 7000rpm for 5 mins. The Supernatant is discarded and the pellet was washed by centrifugation in 99.9% Ethanol. Further washings were given in Ethanol: Water mixtures in ratios of 3:1, 1:1, 1:3 and finally in Water. The concentration of Silica in suspension was estimated by Gravimetric estimation. The suspension was diluted to attain concentrations of 20mg/ml in water and stored at room temperature.

#### Functionalization of Silica spheres with GPTES

Surface Functionalization of Silica Spheres was performed as follows, First, the silica particles were diluted to a concentration of 10mg/ml and washed with a mixture of 3 parts concentrated Sulfuric acid and 1 part 30% Hydrogen peroxide (piranha Solution) for 30 minutes to remove any organic debris and create a high density of hydroxyl groups on the surface of silica particles. The particles were washed by centrifugation several times and suspended in 95% Ethanol followed by thorough sonication to break agglomerates. The absence of agglomeration was confirmed every time by optical microscopy. GPTES was added to a final concentration of 0.5%(v/v) and the reaction was carried out with reflux for 24 hours. after 24 hours the unreacted silane was removed by centrifugated and repeated washings. The surface functionalized silica particles were then stored in either ethanol or water at a concentration of 10mg/ml.

#### Enzyme Functionalization

Epoxide-modified silica particles were used at a concentration of 10mg/ml for the immobilization of Urease on silica particles. Typically, the epoxy-activated particles were washed with water several times and suspended in 0.1M carbonate buffer at pH 9.2. This was followed by thorough sonication to break agglomerates. Urease was dissolved in an equal volume of 0.1M Carbonate buffer at pH 9.2 at 6mg/ml. Both the solutions were mixed with shaking, at this point the concentration of silica particles in the reaction mixture is 5 mg/ml and that of enzyme urease is 3 mg/ml. The reaction was continued with shaking at Room temperature for 24 hours. After 24 hours, the particles were centrifuged and washed with buffer.

Synthesis of Silica stabilized Wax Colloidosomes

The Synthesis of Janus particles was done following the method described by Hong et al[37]. with modifications from Jiang et al.[38] A typical reaction consisted of 20ml of GPTES modified Silica in water with surfactant DDAB and 2 Grams of Paraffin wax. First silica suspension was diluted to 10mg/ml to a final concentration of 20 ml in Water and varying concentrations of DDAB from 15 mg/L to 60mg/L. The solution was sonicated to break agglomerates and then heated to 70°C. 2 grams of Paraffin wax were also separately heated to 70°C to melt. The molten wax was then added to the silica suspension with very high stirring at 2000rpm. The reaction was continued for two hours while maintaining the stirring speed. After 2 hours, 50ml of cold water (4°C) was poured directly into the emulsion to rapidly quench it. This would yield fine colloidosomes which float on the surface of the water.

The Colloidosomes were filtered with an 800nm filter membrane and washed with several volumes of water and ethanol to remove unbound and loosely bound silica, along with DDAB. The washed Colloidosomes were then air-dried and stored at Room Temperature for further use. The Colloidosomes were suspended in 20ml of 0.1M carbonate buffer at pH 9.2. A small amount of Ethanol was also added to the suspension to prevent hydrophobic aggregation of Colloidosomes in water. To this Urease solution was added to a final concentration of 3 mg/ml. The reaction was continued with shaking at Room temperature for 24 hours.

After 24 hours, the Colloidosomes were filtered with an 800nm membrane filter and washed with buffer. The Urease immobilized Colloidosomes were then air-dried. To recover Janus functionalized silica spheres from paraffin Colloidosomes, the dried Colloidosomes were dispersed in 99.99% n-Hexane dissolving the paraffin wax. The now free Silica spheres were collected by centrifugated at 7000rpm for 5 mins. The particles were washed with several volumes of Hexane to remove any undissolved wax and finally suspended in Phosphate buffer.

Particle Size Measurements

The silica colloidal solution was diluted to 0.1mg/ml in water and Particle size analysis was carried out based on DLS in a Malvern Zetasizer Nano ZS. After 3 measurements each consisting of a minimum of 13 runs the mean particle size and PDI was recorded.

Zeta Potential

The Silica colloidal solution was diluted to 0.1mg/ml and the sample was filled in a DTS1070 Folded capillary Zeta cell. The analysis was carried out in Malvern Zetasizer Nano ZS. After 3 measurements with a minimum of 20 runs, the Zeta potential was recorded.

Fluorescence Activated Cell Cytometry (FACS)

All 5 samples including controls were prepared and their concentration was adjusted by turbidometry to $1 \times 10^6$ particles per ml. FACS measurement was done with a BD FACSAria™ III Cell Sorter with a 70-micron nozzle. The FITC intensity histogram of the samples was recorded along with forward and side scatter parameters. The density distribution in the negative control sample was used to create a gate called "silica particles". This gate was applied to all samples and any event outside this gate was excluded from the experiment.

**Acknowledgements**

The research was supported by the National Centre for Nanoscience and Nanotechnology, University of Mumbai. The authors would also like to thank IIT Bombay and the IRCC Flow cytometry facility for assistance with FACS.